\documentclass[10pt,journal,compsoc]{IEEEtran}
\ifCLASSOPTIONcompsoc
  \usepackage[nocompress]{cite}
\else
  \usepackage{cite}
\fi
\usepackage{graphicx}
\usepackage{textcomp}

\usepackage{amsmath}
\usepackage{amssymb}
\usepackage{booktabs}
\usepackage[noend]{algpseudocode}  
\usepackage{booktabs}
\usepackage{multirow}
\usepackage[ruled,vlined,linesnumbered,boxed,commentsnumbered]{algorithm2e}

\ifCLASSINFOpdf
\else
\fi
\begin{document}
%
\title{Robust Split Federated Learning for U-shaped Medical Image Networks}
%
%
%
%

\author{Ziyuan Yang, Yingyu Chen, Huijie Huangfu, Maosong Ran, Hui Wang, Xiaoxiao Li, \IEEEmembership{Member, IEEE}, and Yi Zhang, \IEEEmembership{Senior Member, IEEE}
\IEEEcompsocitemizethanks{\IEEEcompsocthanksitem Z. Yang, Y. Chen, H. Huangfu, M. Ran and H. Wang are with the College of Computer Science, Sichuan University, Chengdu, 610065, China\protect\\
E-mail: cziyuanyang@gmail.com; 2021323040005@stu.scu.edu.cn; huangfu.huijie@qq.com; maosongran@gmail.com; whmzfc@gmail.com
\IEEEcompsocthanksitem X. Li is with the Department of Electrical and Computer Engineering, The University of British Columbia, Vancouver, BC, V6T1Z4, Canada\protect\\
E-mail: xiaoxiao.li@ece.ubc.ca
\IEEEcompsocthanksitem Y. Zhang is with the School of Cyber Science and Engineering, Sichuan University, Chengdu 610065, China\protect\\
E-mail: yzhang@scu.edu.cn
}
\thanks{
Y. Zhang is the corresponding author.}}

\IEEEtitleabstractindextext{%
\begin{abstract}
U-shaped networks are widely used in various medical image tasks, such as segmentation, restoration and reconstruction. Most of them usually rely on centralized learning and thus ignore privacy issues. To address privacy concerns, federated learning (FL) and split learning (SL) have attracted increasing attention. However, it is hard for both FL and SL to balance the local computational cost, model privacy and parallel training simultaneously. To achieve this goal, in this paper, we propose \textbf{Ro}bust \textbf{S}plit \textbf{F}ederated \textbf{L}earning (\textbf{RoS-FL}) for U-shaped medical image networks, which is a novel hybrid learning paradigm of FL and SL. To preserve data privacy, including the input, model parameters, label and output simultaneously, we propose to split the network into three parts hosted by different parties. Besides, distributed learning methods usually suffer from a drift between local and global models caused by data heterogeneity. Based on this consideration, we propose a \textbf{D}ynamic \textbf{W}eight \textbf{C}orrection \textbf{S}trategy (\textbf{DWCS}) to stabilize the training process and avoid model drift. The effectiveness of the proposed RoS-FL is supported by extensive experimental results on different tasks \footnote{Related codes will be released at https://github.com/Zi-YuanYang/RoS-FL.}.
\end{abstract}

\begin{IEEEkeywords}
Split Federated learning, privacy, U-shaped medical image network, dynamic weight correction.
\end{IEEEkeywords}}

\maketitle

\IEEEdisplaynontitleabstractindextext

%
\IEEEpeerreviewmaketitle

\IEEEraisesectionheading{\section{Introduction}\label{sec:introduction}}

%
%
%
%
\IEEEPARstart{I}{n} the past decade, U-shaped medical image networks have achieved great success in various medical image tasks, such as segmentation \cite{Survey_Ushaped}, restoration \cite{Uformer}, reconstruction \cite{xia2022synergizing} etc. These methods require numerous data to train a network model. However, it is difficult to collect a large amount of data in a single institution in realistic scenario. Meanwhile, gathering data from multiple sources is hindered by several data protection regulations, especially for the privacy issues. To address this problem, distributed collaborative machine learning (DCML) has attracted increasing attention \cite{xu2022hercules, liu2021revfrf, xu2020privacy}. It enables training a model in a decentralized manner without accessing the raw data from a single institution \cite{thapa2021advancements}.

As a representative DCML paradigm, Federated Learning (FL) \cite{huang2021fl,jiang2022dynamic} facilitates parallel training and their training time overhead is satisfactory. However, there are two challenges in FL: (1). \textit{The Requirement of Client computational resource}: Each client needs to train a full model, which requires high computational resources for each client. Hence, FL may be not suitable for computational resource-constrained environments. (2). \textit{Model Privacy}: All participants, including clients and the server, have full access to the whole model, which increases the potential privacy and security risks \cite{SFL}.

Recently, Split Learning (SL) was proposed to remedy FL's shortcomings by splitting the full model into server-side and client-side models \cite{SplitLearning}. The splitting operation has two advantages. First, clients only need to train a part of the model, and the most resource-consuming part is trained in the server. Second, each party, including clients and server, only partially accesses the full model. Hence, SL provides better model privacy and requires fewer client-side computational resources than FL. Despite the above merits, the main concern of SL is the training cost. The training process of SL is sequential, which only allows one client to engage with the server at one instance \cite{slcommu}.

Except for the above issues, data heterogeneity is an opening challenge for all DCML paradigms \cite{qu2022rethinking}. This problem leads to a huge gap between the optimization directions of local and global models. For clients, local optimization attempts to find an optimal model for local data with a disinterest in global performance.  global optimization expects to find a globally optimal solution, which has a certain tolerance for sub-optimal local performance. The different optimization goals will cause a drift between local and global models. 

To relieve the above problems in a unified framework, we propose \textbf{Ro}bust \textbf{S}plit \textbf{F}ederated \textbf{L}earning (\textbf{RoS-FL}) for U-shaped medical image networks. RoS-FL is a hybrid learning paradigm, which combines SL and FL and attempts to achieve the best of both worlds. We notice that, unlike classification task, the labels and outputs in segmentation, restoration and reconstruction tasks usually contain privacy information. To protect data privacy and reduce the computational costs in client side, we split the U-shaped network into three parts, including the head, body, and tail models. The proposed framework consists of three parts, which are computation server, aggregation server, and clients, respectively.
The head and tail models are hosted in clients, and the body part which requires large computational resource is hosted in computation server. To enable parallel training, each client corresponds to one body model in computation server. All clients perform the forward propagation of client-side models in parallel and upload the outputs of head models to computation server. 
Body models are executed separately in parallel and transmit their outputs to the clients as the inputs of tail models. 
The loss computation and back-propagation are independently performed in each client, and then the gradients of intermediate results are transferred to update the body and head models. After one communication round, client-side and server-side models are aggregated in the aggregation and computation servers, respectively.
It can be seen that the input, label and output are not transferred to other parties and each party only partially accesses the full model in the whole learning process. As a result, our method can effectively protect data privacy.

In addition, to relieve the data heterogeneity problem, we propose a dynamic weight correction strategy (DWCS) to relieve the model drift problem and correct the global model. Specifically, a weight correction loss is designed to quantify the drift between models from two adjacent communication rounds, and the correction model is optimized by minimizing this loss. Then, we treat the weighted sum of the correction and last round models as the final corrected model. In early training, a large weight will be assigned to the model of current round, since the model is under-fitting. As the training continues, since the local training in the clients may lead to the model drift problem, a large weight for the correction model is necessary to recover from this dilemma and approach the global optimal solution. In summary, we expect to stabilize the training process for a robust model and the weights of the two models are dynamically tuned to avoid model drift. The main contributions of this paper are summarized as:

\begin{itemize}
\item To simultaneously protect the privacy of input, model parameters, label and output, we propose a novel distributed learning framework RoS-FL for U-shaped networks, which can further facilitate parallel training and reduce local computation cost.

\item To relieve the problems of data heterogeneity and model drift, we propose DWCS to dynamically correct the global model.

\item 
Extensive experiments are conducted on medical image segmentation and restoration tasks, and our method demonstrates superior performance to competing methods in both tasks.
\end{itemize}

\section{Related Works}
\label{sec:related works}
\subsection{U-shaped Medical Image Networks}
Since U-Net was proposed \cite{unet}, most segmentation methods chose U-shaped architecture as the backbone and achieved encouraging performances \cite{peng2020method,lei2020skin, lei2020self}.
For example, Fakhry \textit{et al.} \cite{fakhry2016residual} combined the residual connection with U-Net \cite{unet}. Similarly, some researchers used DenseNet block to replace regular convolutional layer \cite{wang2019densely,li2018h,zhou2019unet++}. 
Oktay \textit{et al.} \cite{oktay2018attention} proposed a novel attention gate segmentation model, which is dubbed Attention U-Net.
Isensee \textit{et al.} \cite{isensee2021nnu} proposed nnUNet, which trains the vanilla U-Net with multiple preprocessing steps and surpassed most existing approaches.
The success of these methods demonstrates the effectiveness of U-shaped structure in medical segmentation \cite{unetreview}.

Besides, U-shaped architecture is also widely used in medical restoration and reconstruction tasks \cite{kulathilake2021review}. For example, Chen \textit{et al.} \cite{chen2017low} introduced the residual block into the U-shaped autoencoder for low-dose CT restoration, which is dubbed residual encoder-decoder convolutional neural network (RED-CNN). Wang \textit{et al.} \cite{wang2022domain} proposed generative adversarial networks with dual-domain U-Net-based discriminators for low-dose CT restoration. Wang \textit{et al.} \cite{Uformer} combined the transformer block and U-Net, and achieved impressive performance. On the other hand, many recent proposed works chose U-shaped networks as the benchmark for medical image reconstruction \cite{lee2020mu,huang2021gan,han2018framing}.
Although the above segmentation and imaging methods achieved competitive performance, they need to collect numerous samples from multiple different data sources and usually ignore data privacy.

\subsection{Distributed Collaborative Machine Learning}
FL is one of the most popular DCML paradigms, which trains a full network at each client in parallel, and then the local gradients are transferred to the server for aggregation \cite{li2020federatedchallenge, huang2022learn,chen2022federated }. Typically, McMahan \textit{et al.} \cite{mcmahan2017communication} proposed FedAvg, which learns the global model by aggregating local models. Furthermore, Li \textit{et al.} \cite{li2020federated} proposed FedProx, which can be considered a re-parametrization of FedAvg. Li \textit{et al.} \cite{li2021fedbn} represented a personalized federated learning method FedBN, which alleviates the feature shift using personalized batch normalization (BN) in clients. In \cite{li2021model}, Li \textit{et al.} compared the local presentation and the global presentation to correct local updates. However, these methods require high client-side computation costs and both the client and server will access the full model, which risks privacy leak.

Recently, SL was proposed to alleviate the above problems by splitting a full model into multiple parts \cite{SplitLearning} and each client only needs to train a part of the model. SL seems to be a better choice than FL in computational resource-constrained environment \cite{singh2019detailed}. In \cite{shan2021towards}, graph neural network is combined with SL to protect model privacy. Jeon \textit{et al.} \cite{jeon2020privacy} proposed parallel SL (PSL) learning to reduce the training time overhead. In this method, each client preserves its local model and doesn't upload it to other clients.

Recently, Researchers have been enthusiastic about introducing DCML to the healthcare field to protect data privacy. Feng \textit{et al.} \cite{feng2022specificity} proposed a personalized magnetic resonance imaging method FedMRI, which consists of a globally shared encoder and client-specific decoders. Yang \textit{et al.} \cite{yang2022hypernetwork} leveraged the prior information of scanning parameters to modulate different local models for CT imaging. In \cite{liu2021feddg}, federated domain generalization (FedDG) utilized the frequency information from different clients to handle the data heterogeneity. Park \textit{et al.} \cite{park2021federated} proposed a multi-task federated learning method for COVID-19 segmentation, detection and classification. Poirot \textit{et al.} \cite{poirot2019split} introduced SL for disease classification. Roth \textit{et al.} \cite{roth2022split} combined split learning and U-Net, but labels and inputs are hosted in different parties, which is against the privacy settings.

However, FL and SL-based methods are hampered in some challenges \cite{turina2021federated}. SL has the advantage of model privacy protection, but it cannot achieve parallel training. FL can parallelize training, but it requires huge client computational sources, and no model privacy is committed. To enjoy the best of both worlds, Thapa \textit{et al.} \cite{thapa2022splitfed} proposed split federated learning (SFL), the hybrid of FL and SL, which achieved satisfactory training overhead and prediction accuracy simultaneously. Zhang \textit{et al.} \cite{zhang2022splitavg} evaluated the performance of split federated learning in several medical tasks. Since SFL transfers outputs or labels between different parties, violating the privacy setting, it is unsuitable for the widely used U-shaped networks for medical image segmentation and reconstruction. Besides, its other limitation is it suffers from the data heterogeneity problem.


 

\section{Proposed Method}
\label{sec:proposed}
\subsection{Problem Formulation}

Suppose that there are $N$ clients, denoted as $C_1$, ..., $C_N$. Each client $C_i$ has a local specific dataset $\mathcal{D}_i$. The goal of FL is to learn a full model $\mathcal{M}$ from $\mathcal{D} = \bigcup_{i=1}^N \mathcal{D}_{i}$. Then, the optimization of FL can be formulated as follows:
\begin{equation}
    \underset{\theta}{\arg \min } \mathcal{L}=\sum_{i=1}^{N} \frac{\left|\mathcal{D}_{i}\right|}{|\mathcal{D}|}\mathbb{E}_{(x, y)\sim \mathcal{D}_{i}}\left[\mathcal{L}_{i}(\mathcal{M}(x,\theta), y)\right],
    \label{equ:1}
\end{equation}
where $\mathcal{L}_{i}$ denotes the loss function for $C_i$, and $\mathcal{L}$ means the overall loss function across all clients. $x$ and $y$ represent the input and label, respectively. $|\mathcal{D}_{i}|$ and $|\mathcal{D}|$ are the number of samples in $\mathcal{D}_{i}$ and $\mathcal{D}$. $\theta$ is the parameter set of $\mathcal{M}$.

Different from FL, whose optimal objective is the full network, the optimal objective of SL is composed of multiple partitions. Assume $\mathcal{M}$ is split into two parts, the client-side model $\mathcal{M}_{c}$ and the server-side model $\mathcal{M}_{s}$. The optimization of SL is formulated as:
 
\begin{equation}
    \underset{\theta_{s}, \theta_{c}}{\arg \min } \mathcal{L}=\sum_{i=1}^{N}\mathbb{E}_{(x, y)\sim \mathcal{D}_{i}}\left[\mathcal{L}_{i}(\mathcal{M}_{s}(\mathcal{M}_{c}(x,\theta_c),\theta_s), y)\right],
        \label{equ:2}
\end{equation}
where $\theta_c$ and $\theta_s$ are the parameter sets of $\mathcal{M}_c$ and $\mathcal{M}_s$, respectively.

\begin{figure}
\vspace{0.1cm}
    \centering
    \includegraphics[width=\linewidth]{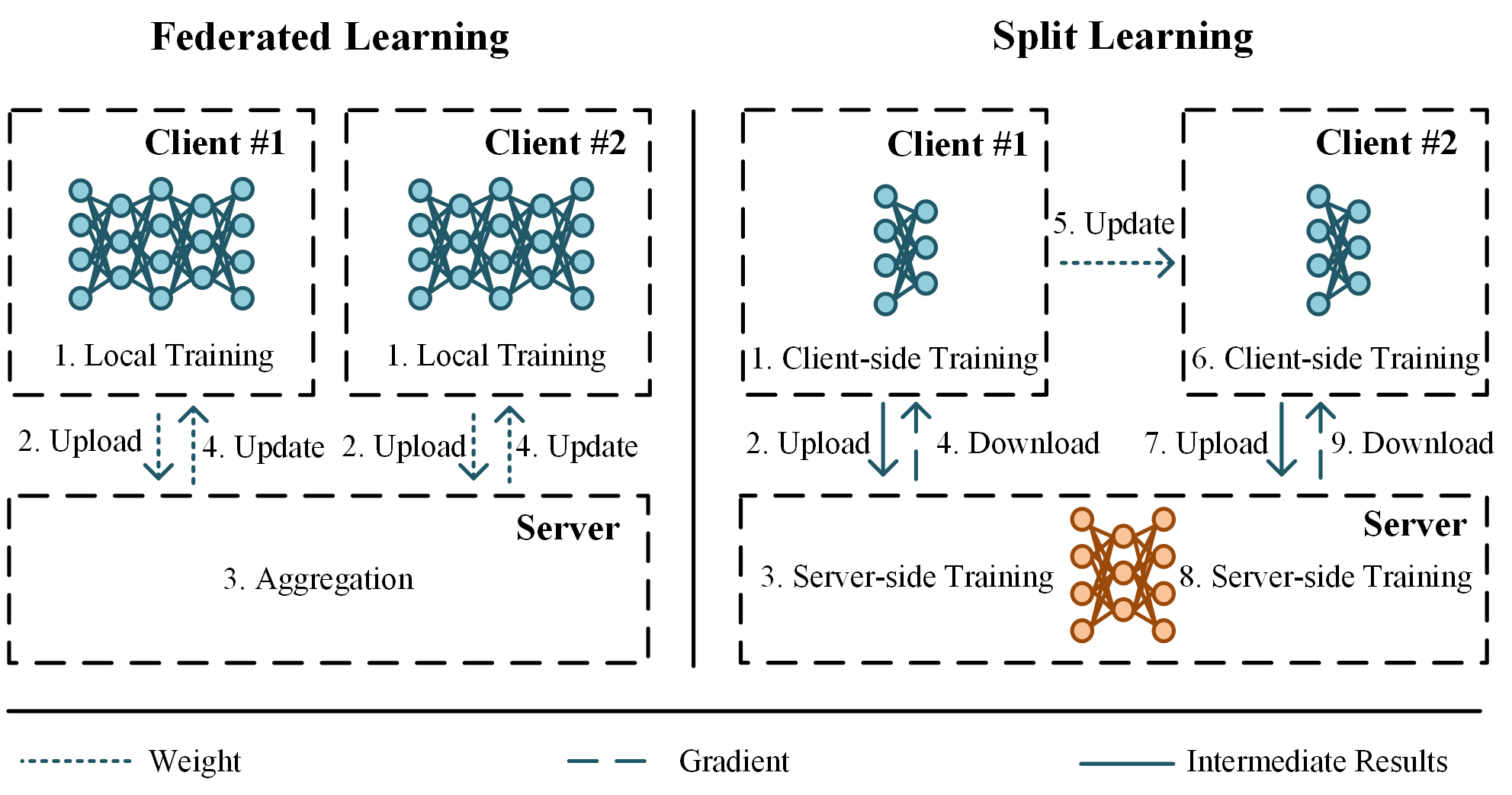}
    \caption{The learning process of FL and SL. The numbers represent the order of processing, and different lines denote different kinds of dataflow.}
    \label{fig_FLandSL}
\end{figure}

\subsection{Architecture of RoS-FL}

\begin{figure*}
  \centering
  
  \begin{minipage}[t]{0.29\linewidth}
  \centering
  \includegraphics[width=\textwidth]{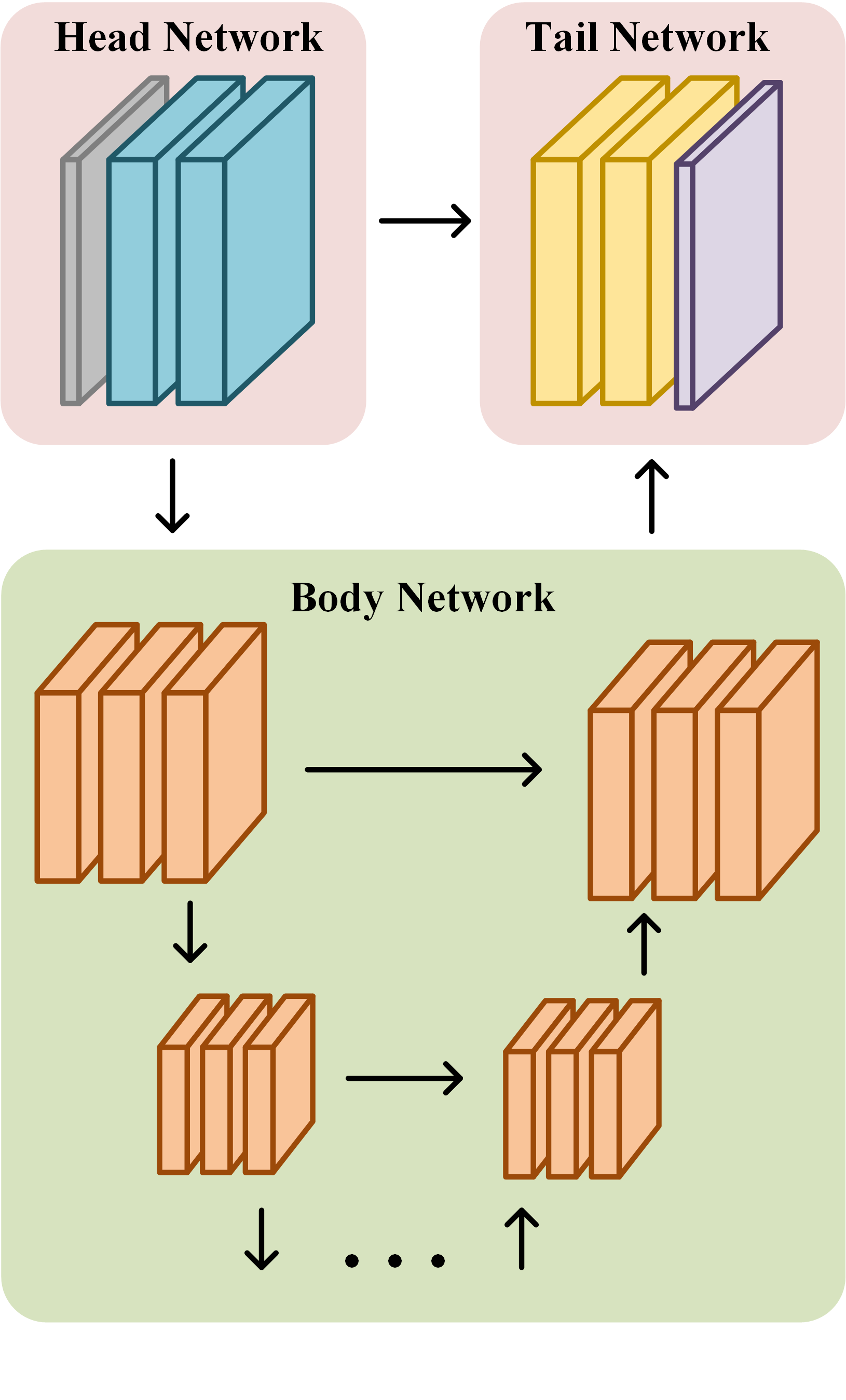}
   \centerline{(a)}
   \label{fig:splitmethod}
   \end{minipage}  
   \quad 
  \begin{minipage}[t]{0.65\linewidth}
  \centering
  \includegraphics[width=\textwidth]{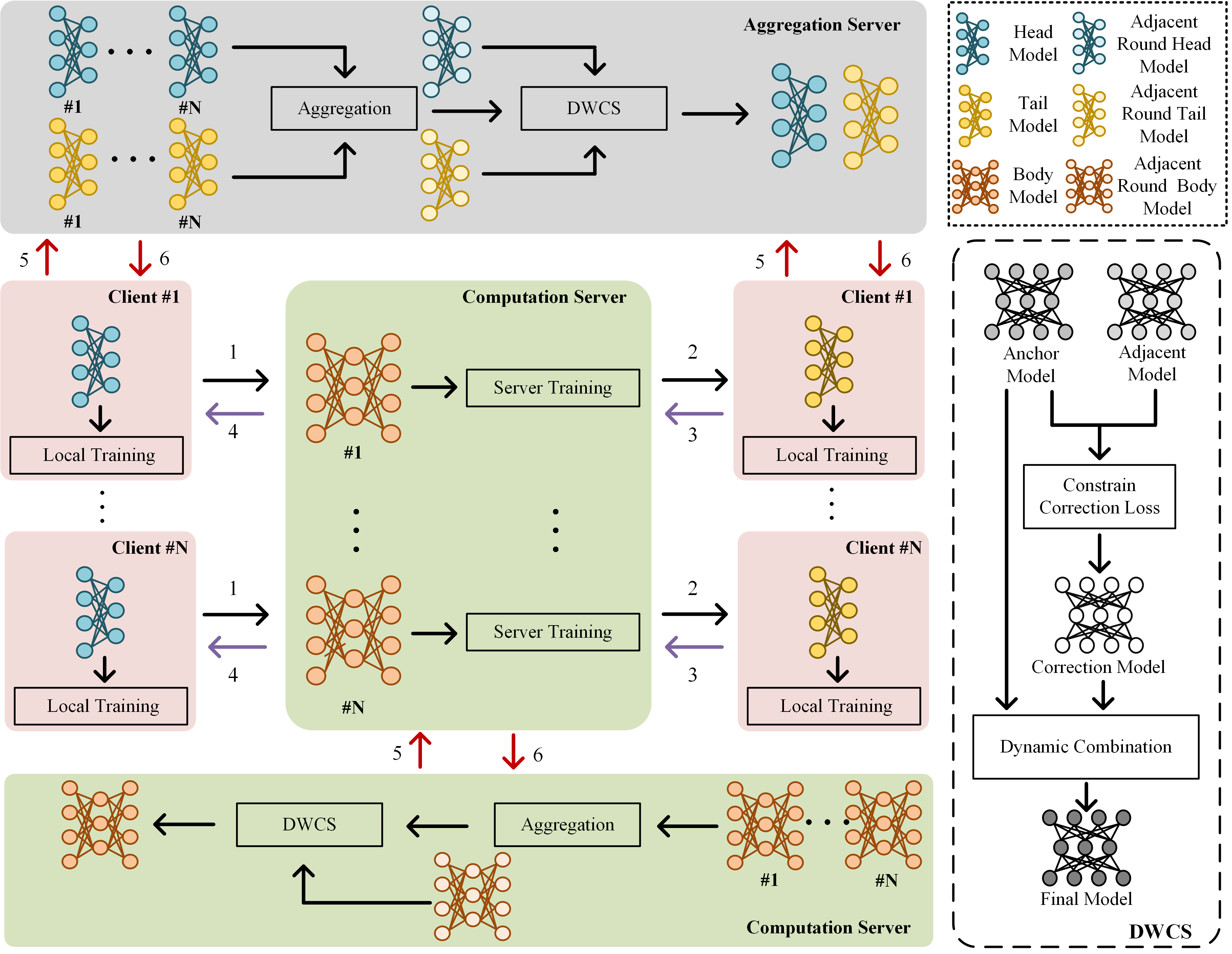}
   \centerline{(b)}
   \label{fig:learnprocess}
   \end{minipage} 
     \caption{The overall of the proposed RoS-FL. Fig. \ref{fig:Framework}(a) represents the proposed split method for U-shaped networks. Fig. \ref{fig:Framework}(b) denotes the whole framework of the proposed RoS-FL, the numbers represent the order of processing. Black arrow, purple arrow, red arrow denote intermediate result, gradient and weight dataflows, respectively.}
  \label{fig:Framework}
\end{figure*}

The learning processes of FL and SL are illustrated in Fig. \ref{fig_FLandSL}. It can be observed that the optimization objectives of these two methods are different. FL expects an optimal full network, but SL tries to optimize the different partition networks. Besides, SL can effectively protect the model privacy, and its local computational costs are low. However, the learning process of SL is sequential, leading to serious training time overhead and under-utilization of client computational resources. Compared with SL, FL can be implemented in a parallel way, but its clients need to train the full models locally, which requires high client-side computational resources.

Actually, SL and FL are complementary and the drawbacks of SL can be fixed by FL and vice-versa. The split operation in SL can reduce the client-side computational cost and protects the model privacy. The aggregation operation in FL can take full advantage of local computational resources and significantly lower the training time overhead. To enjoy the merits of both worlds, we propose a hybrid learning paradigm for U-shaped medical image networks. A similar idea 
was also proposed in \cite{SFL}, but its splitting method is not suitable for segmentation, restoration, and reconstruction. The main reasons lie in that: 

1) \textit{Violating privacy setting}: This method transfers the label or output to other parties and these data contain patients' privacy information.

2) \textit{Extra bandwidth}: SFL requires extra bandwidth to transfer intermediate features of shortcut connections between different parties. However, shortcut connection is the main contribution for U-shaped medical image networks, and it is inappropriate to delete them to improve communication efficiency.

The above problems motivate us to propose a split method without sharing the input, model parameters, output and label of different parties. As shown in Fig. \ref{fig:Framework}(a), the full network is split into three parts, including head, body, and tail networks. The lightweight head and tail networks are hosted in clients to reduce the local computational costs, and the computational resource-required body network is hosted in the server with high-performance computational resources. Then, we can formulate the forward process as follows:
\begin{equation}
    \mathcal{F}(x) = \mathcal{M}_{t}(\mathcal{M}_{b}(\mathcal{M}_{h}(x,\theta_h),\theta_b),\theta_t),
\label{equ:3}
\end{equation}
where $\mathcal{M}_{h}$, $\mathcal{M}_{b}$ and $\mathcal{M}_{t}$ denote the head, body and tail networks with parameter sets $\theta_h$, $\theta_b$ and $\theta_t$, respectively. 

As mentioned above, the output of each encoder layer is the input of the corresponding decoder layer. Assuming that the connected encoder and decoder layers are hosted in different parties, the outputs of encoder layers must be uploaded to other party, which will increase the communication cost. In our method, the corresponding encoder and decoder layers are hosted in the same party, so there is no extra communication bandwidth for feature transfer. Notably, the input, label, and output are preserved locally without any sharing requirements in our method, so the data privacy is well protected.

\subsection{RoS-FL Implementation}

Based on the proposed splitting strategy, we propose a collaborative distributed training framework RoS-FL to achieve parallel training and protect data privacy simultaneously. The flowchart of RoS-FL is illustrated in Fig. \ref{fig:Framework}(b). As mentioned above, the clients and server can access the full model in FL, which is against the principle of model privacy protection. To relieve this problem in FL, we assume that three parties, including clients, computation server, and aggregation server, participate in training. Each of them is authorized to only access part of the model. Clients and aggregation server can access the head and tail models, and the computation server access the body model. Besides, to implement parallel training, $N$ body models are built in the computation server corresponding to the $N$ clients, and the body models of all clients are executed separately to reduce the training time overhead.

At the beginning of training, client-side and server-side models are initialized in the aggregation and computation servers, respectively. All clients perform forward propagation of head models locally: $\hat{y}_{h}=\mathcal{M}_{h}(x,\theta_h)$, and then deliver the encoded results to the computation server. 
Benefiting from the setting of multiple body models, the forward propagation of body models can be executed in parallel: $\hat{y}_{b}=\mathcal{M}_{b}(\hat{y}_{h},\theta_b)$.
At the end of the forward path, $\hat{y}_{b}$ is delivered to the clients to generate the final prediction of tail networks as $\hat{y}=\mathcal{M}_{t}(\hat{y}_{b},\theta_t)$.

\begin{algorithm}[h]
  \caption{Main steps of RoS-FL.}  
  \label{alg:Framwork}
   \textbf{Function Main:} \Comment{Computation Server Executes}\\
    Initialize $\theta_{b}^0$\\
    \For{round $k=1,2,...,R$}{
    \textbf{AggSer}($k-1$)\\
    \For{client $n=1,2,...,N$ \rm{\textbf{in parallel}}}{
    $\theta_{b,n}^k\gets\theta_{b}^{k-1}$
    }
        \For{epoch $i=1,2,...,E$}{
            \For{client $n=1,2,...,N$ \rm{\textbf{in parallel}}}{
            $\hat{y}_{h}\gets$\textbf{HeadForward}($n,k$)\\
            $\hat{y}_{b}\gets\mathcal{M}_{b}(\hat{y}_{h},\theta_{b,n}^k)$\\
            $\frac{\partial\mathcal{L}_n}{\partial\hat{y}_{b}}\gets$
            \textbf{TailMain}($n,\hat{y}_{b}$) \& Backprop\\
            \textbf{HeadBack}($n,\frac{\partial\mathcal{L}_n}{\partial\hat{y}_{b}}$)\\
            $\theta_{b,n}^k\gets\theta_{b,n}^k-\eta\frac{\partial\mathcal{L}_n}{\partial\theta_{b,n}^k}$
            }
        }
    $\theta_b^k\gets\sum_{n=1}^{N}\frac{\left|\mathcal{D}_{n}\right|}{|\mathcal{D}|}\theta_{b,n}^k$\\
    $\theta_{b}^{k}\gets$\textbf{Correct}($\theta_{b}^{k},\theta_{b}^{k-1}$)\\
    }

    \textbf{Function HeadForward}($n$): \Comment{Client $C_n$ Executes}\\
    $x_n\gets$ Sampled input batch from $D_i$\\
    return $\mathcal{M}_{h}(x_n,\theta_{h,n}^k)$
    
    \textbf{Function TailMain}($n,\hat{y}_b$): \Comment{Client $C_n$ Exceutes}\\
    $y_n\gets$ Sampled label batch from $D_i$\\
    $\hat{y}_t\gets\mathcal{M}_t(\hat{y}_b,\theta_{t,n}^k)$\\
    $\mathcal{L}_n\gets$TaskLoss$(\hat{y}_t,y_n)$ \\
    Backprop \& $\theta_{t,n}^k\gets\theta_{t,n}^k-\eta\frac{\partial\mathcal{L}_n}{\partial\theta_{t,n}^k}$\\
    return $\frac{\partial\mathcal{L}_n}{\partial\hat{y}_{b,n}^k}$
    \\
    \textbf{Function HeadBack}($n$,$ \frac{\partial\mathcal{L}_n}{\partial\hat{y}_{b}}$): \Comment{Client $C_n$ Executes}\\
    Backprop \& $\theta_{h,n}^k\gets\theta_{h,n}^k-\eta\frac{\partial\mathcal{L}_n}{\partial\theta_{h,n}^k}$\\
    \textbf{Function AggSer}($k$):\Comment{Aggregation Server Exceutes}\\
    \If{$k=0$}{
            Initlialize $\theta_h^k,\theta_t^k$
            }\Else{
                $(\theta_h^k,\theta_t^k)\gets(\sum_{n=1}^{N}\frac{\left|\mathcal{D}_{i}\right|}{|\mathcal{D}|}\theta_{h,n}^k,\sum_{n=1}^{N}\frac{\left|\mathcal{D}_{i}\right|}{|\mathcal{D}|}\theta_{t,n}^k)$\\
                $(\theta_{h}^{k},\theta_{t}^{k})\gets$(\textbf{Correct}($\theta_{h}^{k},\theta_{h}^{k-1}$),\textbf{Correct}($\theta_{t}^{k},\theta_{t}^{k-1}$))\\
            }
            Deliver $\theta_h^k,\theta_t^k$ to clients

        \textbf{Function Correct}($\theta^k,\theta^{k-1}$):\\
        $\mathcal{L}_{con}\gets$WeightCorrectionLoss$(\theta^k,\theta^{k-1})$\\
        $\theta_c^k\gets\theta^k-\eta\frac{\partial\mathcal{L}_{con}}{\partial\theta^{k-1}}$\\
        $\theta_r^k \gets (1-\alpha)\theta^k+\alpha\theta_{c}^k$\\
        return $\theta^k$
\end{algorithm}

After forward propagation, each client calculates the loss and starts backpropagation. Concretely, the gradients about $\mathcal{M}_{t}$ and $\hat{y}_{b}$ are calculated at first. Then, the gradients of $\hat{y}_{b}$ are transmitted to the computation server, and the server executes the backpropagation on $\mathcal{M}_{b}$ and deliver the gradients of $\hat{y}_{h}$ to clients. Finally, with the received gradients of $\hat{y}_{h}$, the client executes the backpropagation of $\mathcal{M}_{h}$. So far, one backpropagation pass between clients and the server is completed. To make full use of all local data and get optimal global models, we aggregate the client-side and server-side models in the aggregation and computation servers, respectively. Until now, one complete global training round has been finished. It can be observed that the training processes of different clients are executed in parallel, which greatly reduce the training time overhead of SL. Meanwhile, there is no party in our framework can access the full model, which effectively protects model privacy. Then, the servers provide aggregated models by averaging local models as:

\begin{equation}
\theta^k=\sum_{i=1}^{N}\frac{\left|\mathcal{D}_{i}\right|}{|\mathcal{D}|}\theta_{i}^k,
    \label{equ:4}
\end{equation}
where $k$ and $i$ represent the current training round and client index, respectively. $\theta^k=\{\theta^k_{h},\theta^k_{b},\theta^k_{t}\}$ represents the parameter sets of $\mathcal{M}^{k}=\{\mathcal{M}_{h}^{k},\mathcal{M}_{b}^{k},\mathcal{M}_{t}^{k}\}$. Then our optimization problem is formulated as:
\begin{equation}
        \underset{\theta_{h}, \theta_{b},\theta_{t}}{\arg\min } \mathcal{L}=\sum_{i=1}^{N}\frac{\left|\mathcal{D}_{i}\right|}{|\mathcal{D}|}\mathbb{E}_{(x, y)\sim \mathcal{D}_{i}}\left[\mathcal{L}_{i}(\mathcal{F}(x),y)\right].
        \label{equ:5}
\end{equation}

\subsection{Dynamic Weight Correction Strategy}

The original DCML suffers from the data heterogeneity problem, which leads to a huge gap between the optimization directions of local and global models. Since there is always a distribution gap between $D_i$ and $D$ in practice, local training will lead the local model to work badly in other data domains. As a result, it may generate a poor optimization solution to the global model and cause model collapse after aggregation. This problem happens more commonly in healthcare tasks, in which the collected data inevitably suffer from serious data heterogeneity caused by several factors, such as different hardware, scanning protocols and patients.

To recover from this situation, we propose DWCS to avoid the model drift problem. Specifically, we treat the model of the last communication round as the anchor model and propose a weight correction loss to quantify the drift between the anchor model and its adjacent communication round model. Then we get the correction model by minimizing the weight correction loss, and the weighted sum of the correction and last round models is treated as the final result. The weight correction loss function is defined as:

\begin{equation}
    \underset{\theta^k}{\arg\min }\mathcal{L}_{con}(\theta^k,\theta^{k-1}) = \frac{\mu}{2}||\theta^k-\theta^{k-1}||_2^2,
    \label{equ:6}
\end{equation}
where $\mathcal{L}_{con}$ represents the weight correction loss, and $ \mu$ is the hyperparameter constraining the optimization step factor. Then, our correction model can be formulated as:

\begin{equation}
    \theta_{c}^k = \theta^k + \eta\bigtriangledown\mathcal{L}_{con}(\theta^k,\theta^{k-1}),
    \label{equ:7}
\end{equation}
where $\eta$ is the learning rate, and our correction model is $\theta_{c}^k=\{\theta_{c,h}^k,\theta_{c,b}^k,\theta_{c,t}^k\}$.

In the early stage of training, a small weight should be assigned to the correction model to accelerate convergence. With the training continuing, the model almost converges, but the local training may cause severe model drift, which leads to global model collapse by aggregating. To alleviate the above issue, inspired by \cite{tarvainen2017mean}, we propose a dynamic adjustment strategy to stabilize the training process and minish the model drift. Then, a robust model $\theta_{r}^{k}$ can be obtained with the weighted summation of $\theta_{c}^k$ and $\theta^k$, which can be defined as:
\begin{equation}
    \theta_{r}^k = (1-\alpha)\theta^k+\alpha\theta_{c}^k.
    \label{equ:8}
\end{equation}
where $\alpha=\min(1-\frac{1}{k+1},\beta)$ is the balancing factor, and $\beta$ denotes the maximum constraint value, which is set to $0.99$ in this paper. 

To help readers to follow our method, the main steps of RoS-FL are introduced in Algorithm \ref{alg:Framwork} in pseudocode style. It can be noticed that the proposed RoS-FL is a parallel DCML method without sharing the input, model parameters, output and label. As a result, the model privacy can be well protected in the whole training phase. Additionally, we only need to perform DWCS once during one communication round, so our method is lighter than regularization-based methods.

\begin{table*}[]
\centering
  \caption{Quantitative Results for the Segmentation Task.}
\normalsize
\begin{tabular}{c|c|c|c|c}
\hline
\makebox[0.2\textwidth][c]{Method}          & \makebox[0.15\textwidth][c]{DSC$\uparrow$} & \makebox[0.15\textwidth][c]{HD95$\downarrow$} & \makebox[0.15\textwidth][c]{ASD$\downarrow$} & \makebox[0.15\textwidth][c]{JC$\uparrow$} \\ \hline
CL            & 0.8910 & 2.587 & 0.7169 & 0.8089 \\ \hline
SL \cite{SplitLearning}            & 0.8754 & 3.624 & 1.0272 & 0.7886 \\
PSL \cite{jeon2020privacy}           & 0.8469 & 3.884 & 1.1564 & 0.7461 \\
FedAvg \cite{mcmahan2017communication}       & 0.8493 & 3.747 & 1.1164 & 0.7489 \\
FedProx \cite{li2020federated}       & 0.8692  & 3.044 & 0.9267 & 0.7777 \\
FedBN \cite{li2021fedbn}         & 0.8510 & 4.268 & 1.3180 & 0.7514 \\
FedMRI \cite{feng2022specificity} & 0.4921 & 21.410 & 6.8436 & 0.3583 \\
FedDG \cite{liu2021feddg}        & 0.8704 & 3.419 & 0.9860 & 0.7801 \\ \hline
RoS-FL w/o DWCS & 0.8743 & 3.450 & 1.0302 & 0.7861
\\ 
RoS-FL w/ DWCS  & \textbf{0.8799} & \textbf{2.718} & \textbf{0.8572} & \textbf{0.7932} \\ \hline
\end{tabular}\vspace{0.2cm}
  \label{tab:SegResult}
\end{table*}

\section{Experiments}
\label{sec:exper}
\subsection{Implementation Details}

\begin{figure*}
    \centering
    \includegraphics[width=1\linewidth]{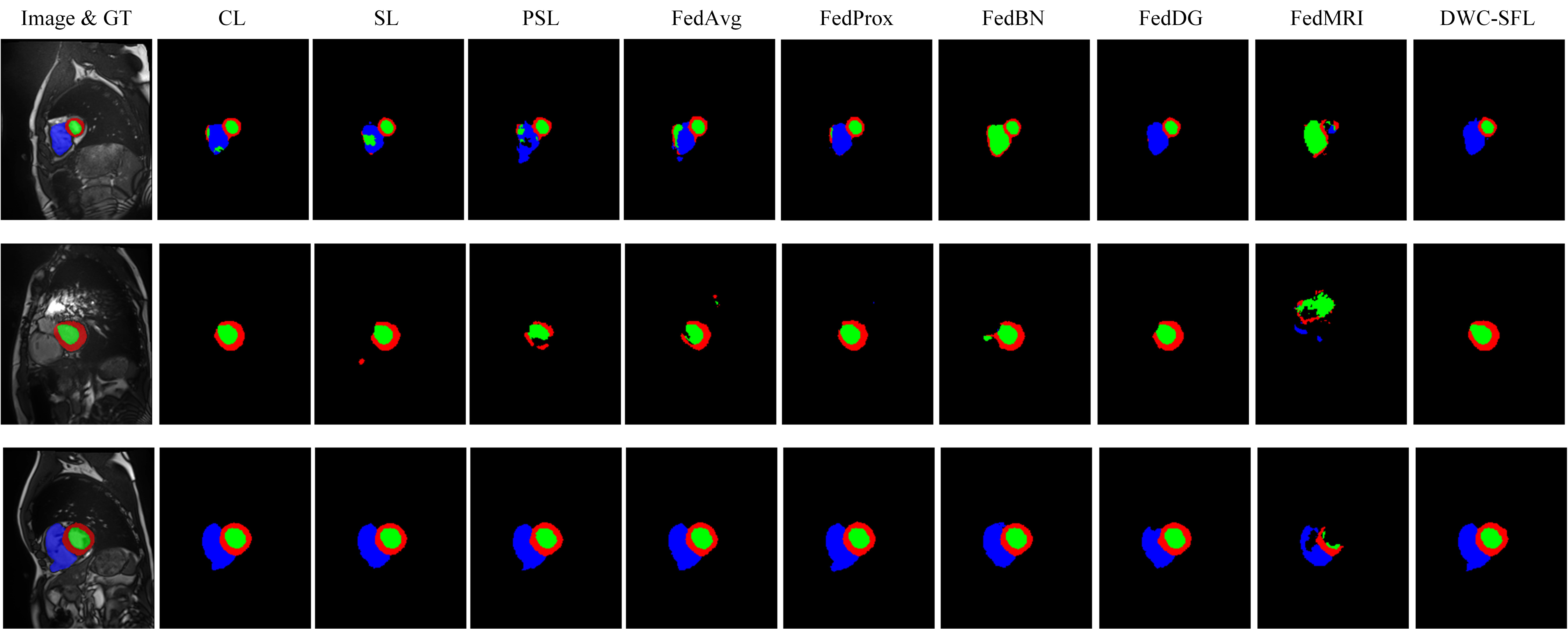}
    \caption{Qualitative comparison on some typical segmentation results of different methods.}
    \label{fig:segresult}
\end{figure*}

To validate the effectiveness of our method for U-shaped medical image networks in different tasks, we compare our method with several state-of-the-art DCML methods, such as SL\cite{SplitLearning}, PSL\cite{jeon2020privacy}, FedAvg\cite{mcmahan2017communication}, FedProx\cite{li2020federated}, FedBN\cite{li2021fedbn}, FedDG\cite{liu2021feddg}, FedMRI \cite{feng2022specificity} and centralized learning (CL) method, which is treated as the benchmark. Adam \cite{kingma2014adam} is adopted to optimize the models. The learning rate is set to $1\times10^{-4}$, and the weight decay is $1\times10^{-8}$. All codes are implemented in PyTorch and the experiments are performed on an NVIDIA GTX 3090 GPU.

\subsection{Segmentation Experiments}
We conduct segmentation experiments on the public dataset Automated Cardiac Diagnosis Challenge (ACDC) \cite{bernard2018deep}, which contains 200 annotated short-axis cardiac MR-cine images from 100 patients. All short-axis slices within 3D scans are resized to 256 $\times$ 256 as 2D images. We randomly divide 80 patients on average into 4 clients as the training set, and the remaining 20 patients are treated as the testing set. Dice Similarity Coeffcient (DSC), 95\% Hausdorff Distance (HD95), Average Surface Distance (ASD), and Jaccard Index (JC) are chosen as the quantitative metrics in this paper. Segmentation networks are all optimized with cross-entropy loss and dice loss. The numbers of communication rounds and local training epochs are set to 300 and 1, respectively.


\begin{table*}[]
\vspace{0.1cm}
\caption{The Geometry Parameters and Dose Levels in Different Clients.}
\centering
\normalsize
\begin{tabular}{c|c|c|c|c}
\hline
             & Client \#1 & Client \#2 & Client \#3 & Client \#4 \\ \hline
Number of views            & 1024 & 128 & 512 & 384 \\
Number of detector bins            & 512 & 768 & 768 & 600 \\
Pixel length (mm)           & 0.66 & 0.78 & 1.0 & 1.4 \\
Detector bin length (mm)        & 0.72 & 0.58 & 1.23 & 1.64 \\
Distance between the source and rotation center (mm)       & 250 & 350 & 500 & 350 \\
Distance between the detector and rotation center (mm)         & 250 & 300 & 400 & 300 \\ 
Intensity of X-rays & 1e5 & 1e6 & 5e4 & 1.25e5\\
\hline
\end{tabular}
\label{tab:geo}
\vspace{0.2cm}
\end{table*}

\begin{figure*}
    \centering
    \begin{minipage}[t]{.74\linewidth}
    \centering
    \includegraphics[width=1\linewidth]{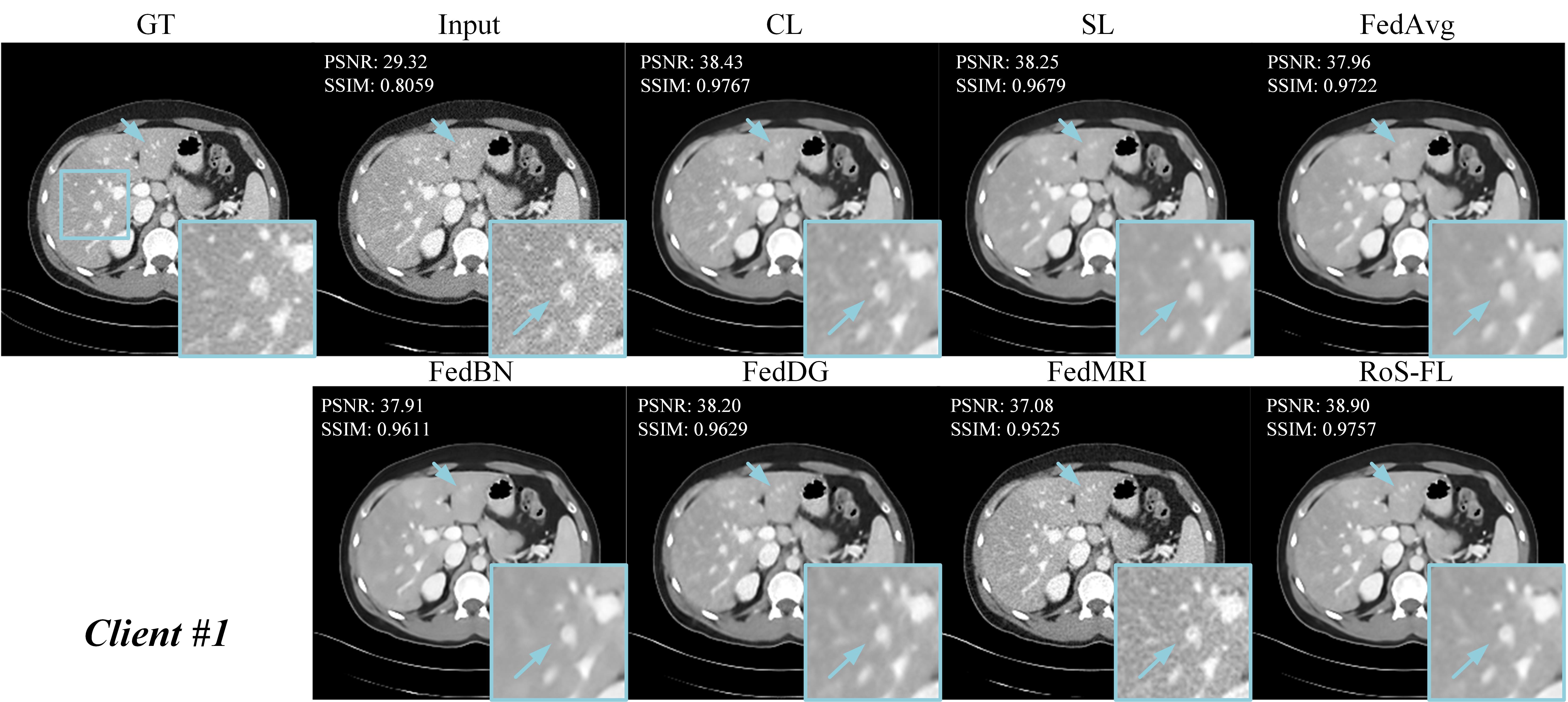}
    \end{minipage}
     \begin{minipage}[t]{.74\linewidth}
    \centering
    \includegraphics[width=1\linewidth]{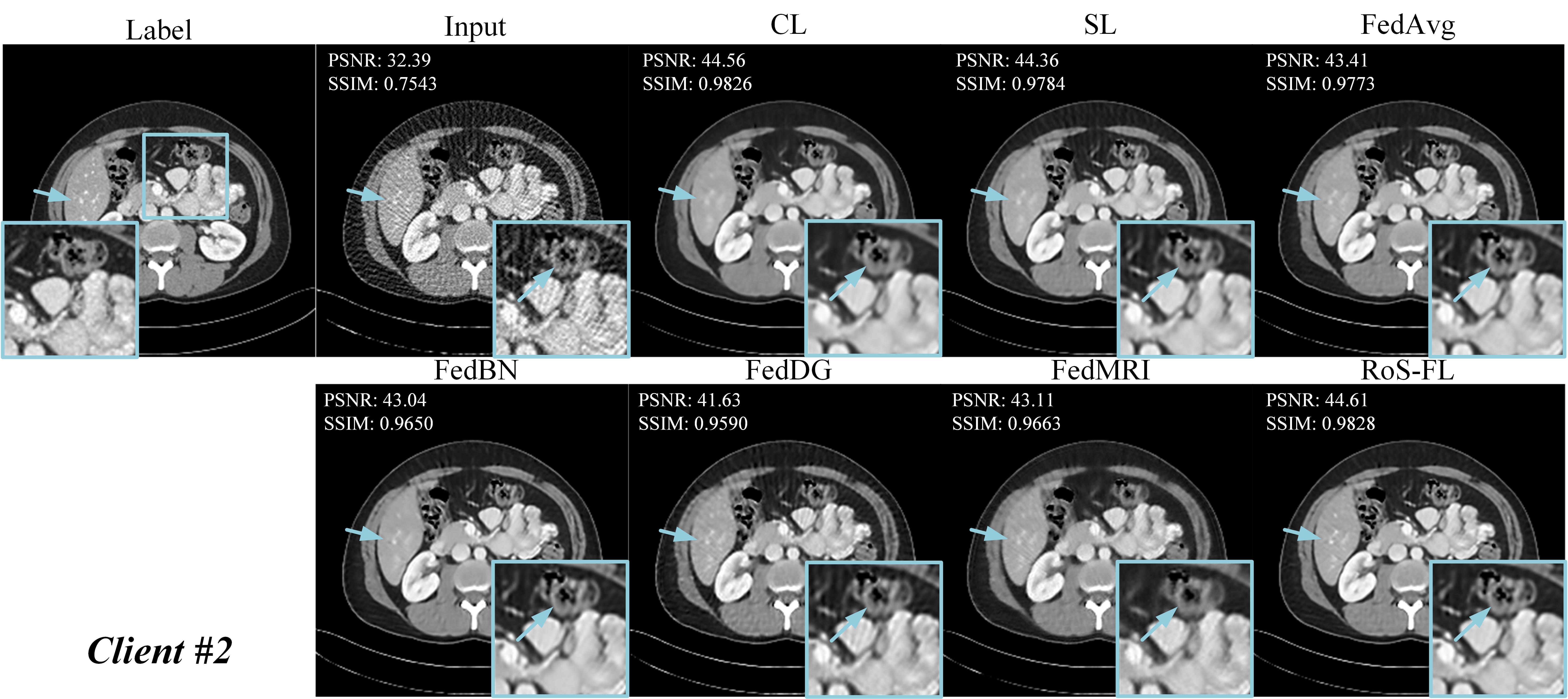}
    \end{minipage}   
        \begin{minipage}[t]{.74\linewidth}
    \centering
    \includegraphics[width=1\linewidth]{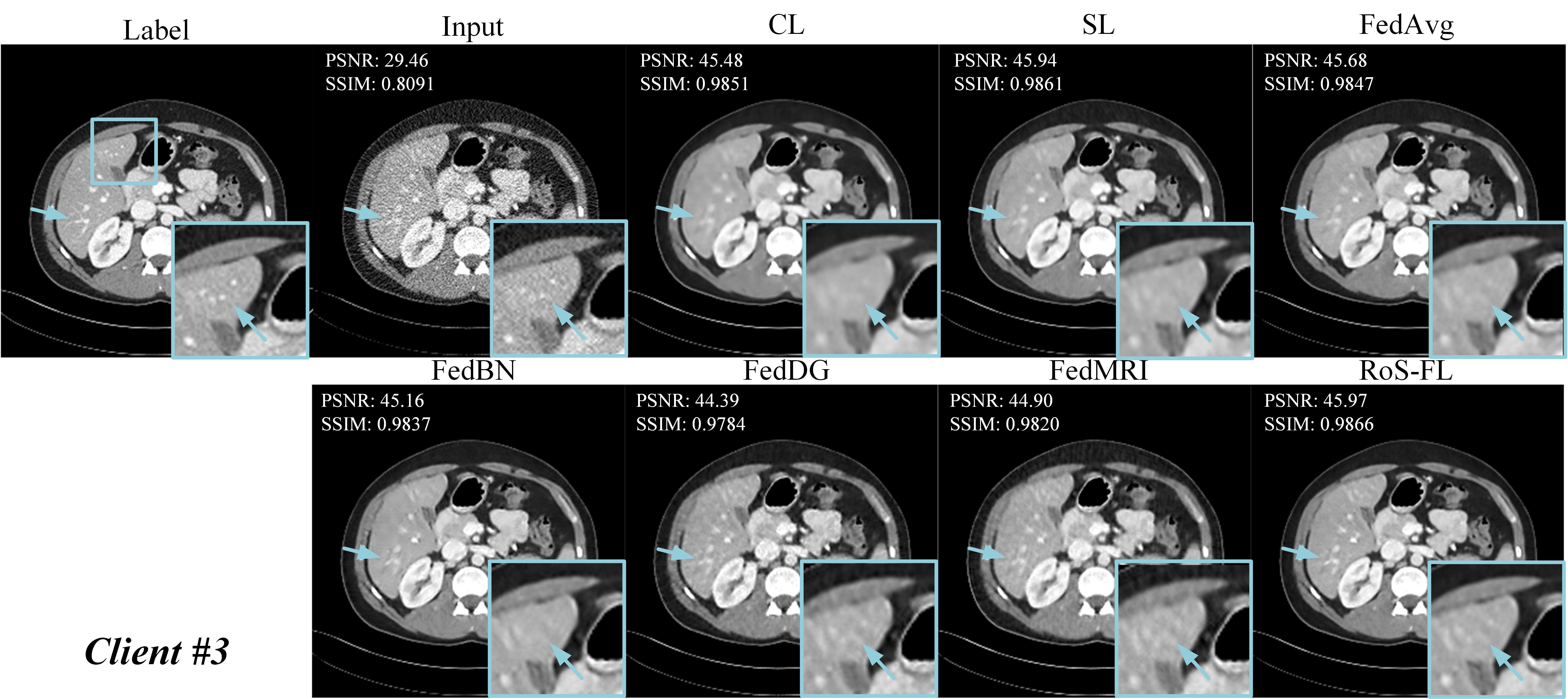}
    \end{minipage}
        \begin{minipage}[t]{.74\linewidth}
    \centering
    \includegraphics[width=1\linewidth]{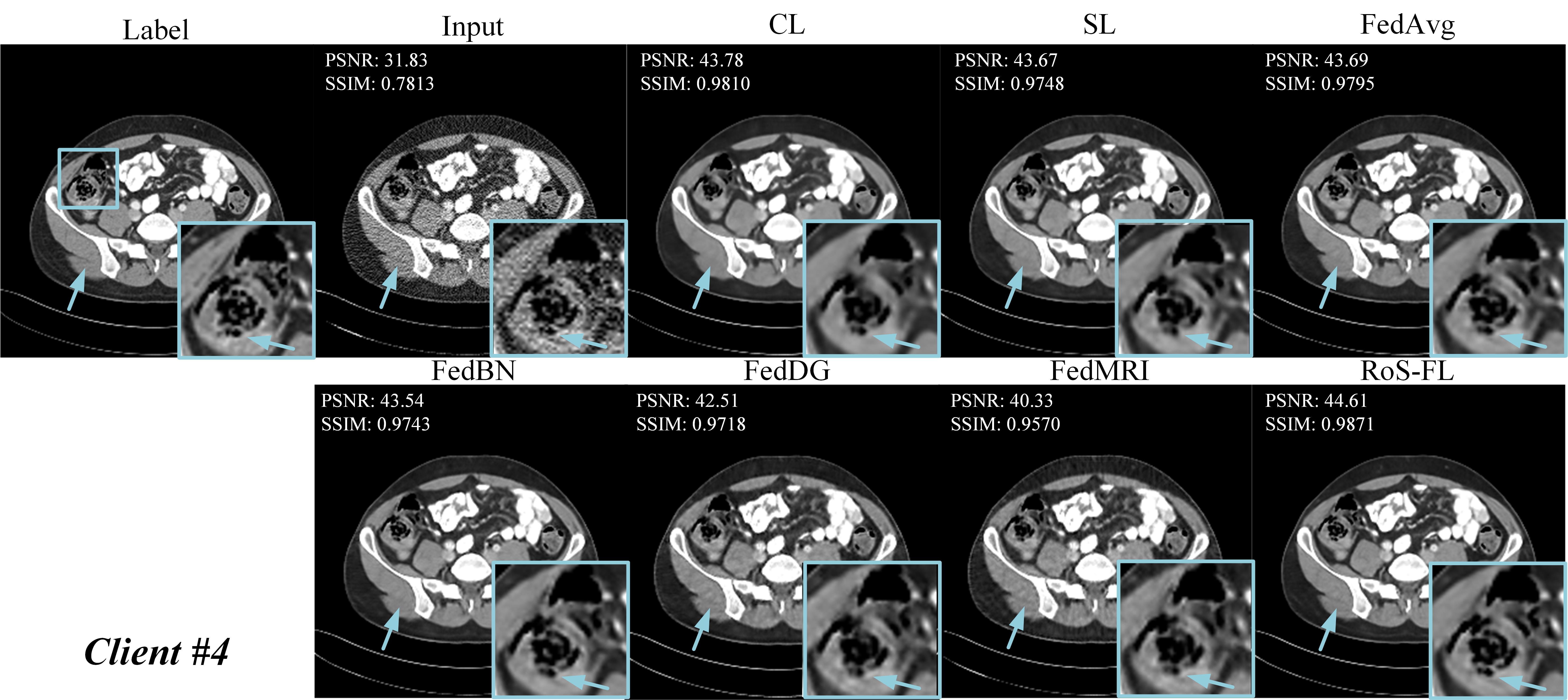}
    \end{minipage}
    \caption{Visual comparisons with state-of-the-art methods on different geometries and dose levels. The display window is [-160,240]HU.}
    \label{fig:imaging_result}
\end{figure*}

The quantitative results are listed in Tab. \ref{tab:SegResult}. It is observed that our method can achieve competitive performance to CL and outperforms other DCML methods. SL achieves satisfactory performance, but its training process is sequential, which leads the training time overhead is $N$ times (the number of clients) longer than our RoS-FL and other FL methods. SL beats most FL-based methods. Our method inherits the advantages of SL and performs much better than FL-based methods. Three representative visual results are shown in Fig. \ref{fig:segresult}. Our method accurately identifies the boundaries of organ regions, and significantly outperforms other methods. In the first case, our result is even better than CL. We must mention that FedMRI is proposed for MRI reconstruction, which is unsuitable for segmentation task.
In the ablation experiment, we find that DWCS can effectively improve the segmentation performance due to its good trade-off between the convergence acceleration and model drift recovery.

\subsection{Restoration Experiments}

\begin{table*}[]
\normalsize
\centering
  \caption{Quantitative Results for the Imaging Task.}
  \vspace{0.05cm}
\begin{tabular}{c|cc|cc|cc|cc|cc}
\hline
              & \multicolumn{2}{c|}{Client \#1}  & \multicolumn{2}{c|}{Client \#2}  & \multicolumn{2}{c|}{Client \#3}  & \multicolumn{2}{c|}{Client \#4} & \multicolumn{2}{c}{Average}   \\
Method        & \multicolumn{1}{c}{PSNR$\uparrow$} & SSIM$\uparrow$ & \multicolumn{1}{c}{PSNR$\uparrow$} & SSIM$\uparrow$ & \multicolumn{1}{c}{PSNR$\uparrow$} & SSIM$\uparrow$ & \multicolumn{1}{c}{PSNR$\uparrow$} & SSIM$\uparrow$ & \multicolumn{1}{c}{PSNR$\uparrow$} & SSIM$\uparrow$ \\ \hline
CL            & \multicolumn{1}{c}{39.35}     &   0.9647   & \multicolumn{1}{c}{42.31}     &  0.9617    & \multicolumn{1}{c}{45.55}     &    0.9859  & \multicolumn{1}{c}{42.83}     &      0.9683 & \multicolumn{1}{c}{42.60}     &  0.9711    \\ \hline
SL \cite{SplitLearning}            & \multicolumn{1}{c}{39.14}     & 0.9562     & \multicolumn{1}{c}{\textbf{41.71}}     & \textbf{0.9592}     & \multicolumn{1}{c}{45.43}     & 0.9863     & \multicolumn{1}{c}{42.79}     & 0.9664     & \multicolumn{1}{c}{42.27}     &   0.9670   \\
PSL \cite{jeon2020privacy}           & \multicolumn{1}{c}{38.75}     & 0.9572     & \multicolumn{1}{c}{41.47}     & 0.9565     & \multicolumn{1}{c}{44.61}     & 0.9831     & \multicolumn{1}{c}{41.58}     & 0.9582     & \multicolumn{1}{c}{41.60}     & 0.9637     \\
FedAvg \cite{mcmahan2017communication}       & \multicolumn{1}{c}{39.67}     & 0.9643     & \multicolumn{1}{c}{40.29}     & 0.9444     & \multicolumn{1}{c}{45.22}     & 0.9851     & \multicolumn{1}{c}{42.73}     &  0.9675    & \multicolumn{1}{c}{41.98}     & 0.9653     \\
FedProx \cite{li2020federated}       & \multicolumn{1}{c}{37.35}     & 0.9309     & \multicolumn{1}{c}{38.10}     & 0.9049     & \multicolumn{1}{c}{43.23}     & 0.9750     & \multicolumn{1}{c}{40.34}     &  0.9471    & \multicolumn{1}{c}{39.75}     & 0.9395     \\
FedBN \cite{li2021fedbn}         & \multicolumn{1}{c}{38.09}     &  0.9477    & \multicolumn{1}{c}{40.01}     &   0.9321   & \multicolumn{1}{c}{44.73}     & 0.9843     & \multicolumn{1}{c}{42.33}     &   0.9628   & \multicolumn{1}{c}{41.29}     &   0.9567   \\
FedDG \cite{liu2021feddg}        & \multicolumn{1}{c}{38.95}     & 0.9528     & \multicolumn{1}{c}{39.34}     & 0.9231     & \multicolumn{1}{c}{44.06}     & 0.9788     & \multicolumn{1}{c}{42.10}     & 0.9623     & \multicolumn{1}{c}{41.11}     & 0.9542     \\
FedMRI \cite{feng2022specificity}        & \multicolumn{1}{c}{37.48}     & 0.9384     & \multicolumn{1}{c}{40.91}     & 0.9451     & \multicolumn{1}{c}{44.57}     & 0.9842     & \multicolumn{1}{c}{39.95}     & 0.9430     & \multicolumn{1}{c}{40.73}     & 0.9524     \\ \hline
RoS-FL w/o DWCS & \multicolumn{1}{c}{39.58}     &  0.9668    & \multicolumn{1}{c}{40.82}     & 0.9527     & \multicolumn{1}{c}{45.17}     &    0.9851  & \multicolumn{1}{c}{43.00}     & 0.9694     & \multicolumn{1}{c}{42.14}     &    0.9685  \\
RoS-FL w/ DWCS  & \multicolumn{1}{c}{\textbf{42.15}}     &    \textbf{0.9749}  & \multicolumn{1}{c}{40.92}     & 0.9513     & \multicolumn{1}{c}{\textbf{45.48}}     &    \textbf{0.9870}  & \multicolumn{1}{c}{\textbf{43.71}}     & \textbf{0.9750}     & \multicolumn{1}{c}{\textbf{42.82}}     &    \textbf{0.9721}  \\ \hline
\end{tabular}\vspace{0.2cm}

  \label{tab:ImgResult}

\end{table*}

The well-known NIH-AAPM-Mayo Low-Dose CT dataset \cite{mccollough2016tu}, which contains 5936 full-dose CT images from 10 patients, is used to verify the performance of our method. Eight patients are randomly divided into four clients on average as the training dataset, and two patients are treated as testing dataset. To simulate real environments, we generate multi-source non-iid low-dose CT (LDCT) data following \cite{xia2021ct, niu2014sparse}. Poisson noise and electronic noise were added to the measured projection data to simulate the low-dose case as follows: 

\begin{equation}
    p=\ln \frac{I_{0}}{\operatorname{Poisson}\left(I_{0} \exp (-\hat{p})\right)+\operatorname{Normal}\left(0, \sigma_{e}^{2}\right)},
\end{equation}
where $\hat{p}$ represents the clean projection, and $\sigma_e$ denotes the variance of electronic noise. $I_0$ represents the number of photons. In this paper, we fixed the electronic noise variance at $\sigma_{e}^{2}=10$ and treat $I_0 = 1\times10^{6}$ following \cite{xia2021ct, niu2014sparse}.

In this paper, four cases with different sparse-view and low-dose data are simulated and the corresponding geometric parameters and dose levels are listed in Tab. \ref{tab:geo}.  Peak Signal to Noise Ratio (PSNR) and Structural Similarity (SSIM) are employed as the quantitative metrics. Restoration networks are optimized with mean-squared error (MSE) loss. The numbers of communication rounds and local training epochs are set to 500 and 1, respectively. 

The quantitative results are shown in Tab. \ref{tab:ImgResult}. It can be noticed that our method achieves the best performance in comparison with other DCML methods, and even works better than CL in some clients. Similar to the segmentation task, FL-based methods are inferior to SL-based methods in most cases. The possible reason lies in that the transferred weights are only with limited information and they cannot make full use of the information from local data. However, as mentioned above, the training time overheads of FL-based methods are much less than those of SL-based methods. Our method combines the merits of both learning paradigms and achieves the best performance. In the ablation experiment about DWCS, we can easily find that it improves the overall performance. We can notice that the results of other methods in Client \#1 have a significant performance gap with those in other clients, which is probably caused by the model drift problem. Benefiting from DWCS, which effectively alleviates the model drift problem by correcting the optimal solution, our method has no noticeable performance gap between different nodes. Fig. \ref{fig:imaging_result} shows several typical slices denoised using different methods. It can be observed that results denoised by other methods still contain noise or artifacts to varying degrees, but our method can effectively remove them. In some results of other methods, edges are blurry and some tiny structures are wrongly restored. Compared with them, our method correctly restores those structural details and edges, and they are clearer in our results.

\begin{table}[]
\centering
\normalsize
\caption{Analysis of Different Numbers of Local Epochs in Restoration Task. (Communication Round/Local Epoch)}
\begin{tabular}{ccccc}
\hline
     & 1/500  & 2/250  & 4/125  & 5/100  \\ \hline
PSNR & 42.81  & 42.38  & 41.89  & 41.90  \\
SSIM & 0.9721 & 0.9699 & 0.9650 & 0.9649 \\ \hline
\end{tabular}
\vspace{0.2cm}
\label{tab:com_restor}
\end{table}

\begin{figure}
  \centering
    \includegraphics[width=\linewidth]{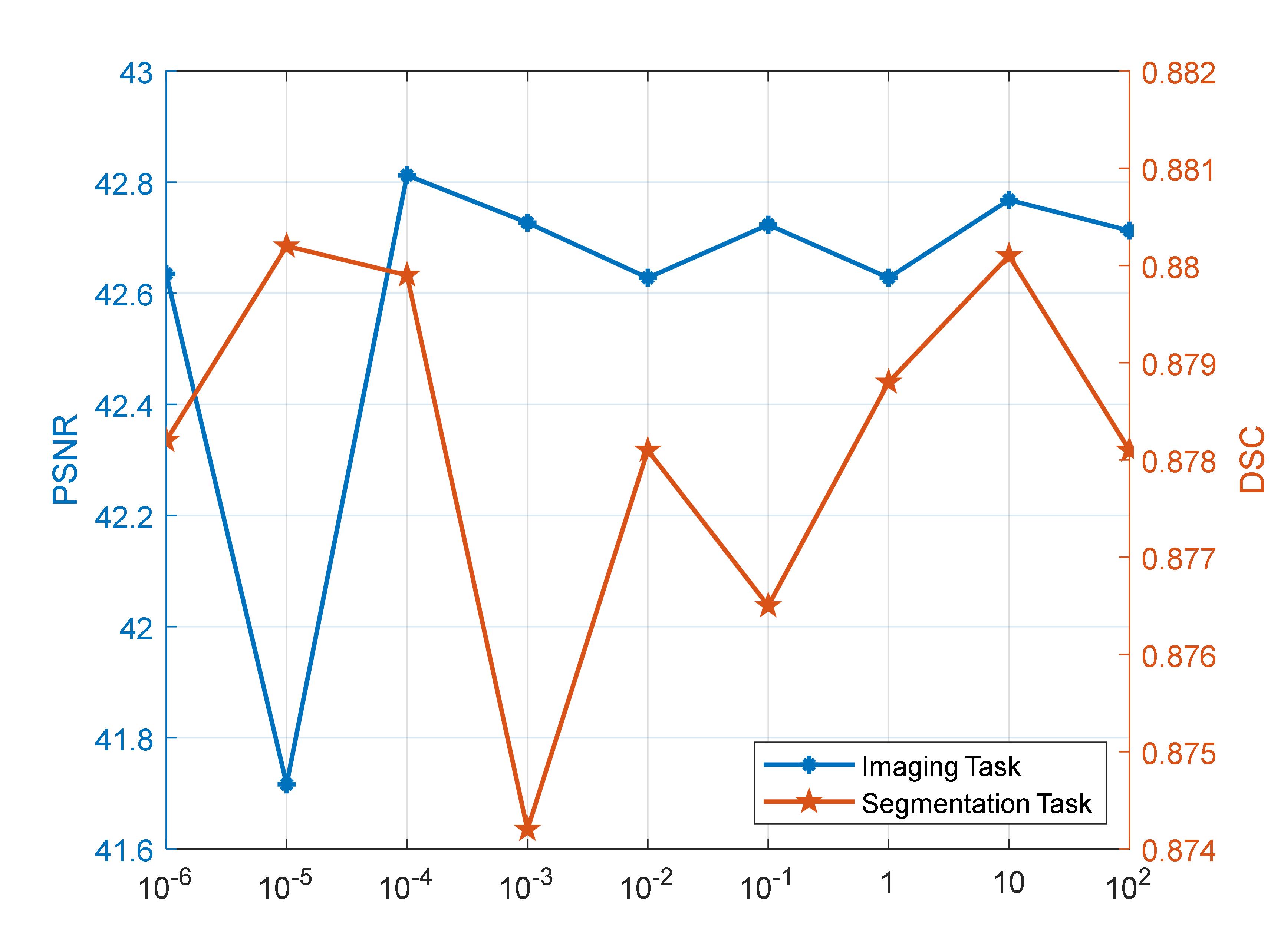}
  \caption{The ablation study about the selection of hyperparameter $\mu$ in Eq. \ref{equ:6}.}
  \label{fig_Ablation}
  \vspace{0.1cm}
\end{figure}

\subsection{Ablation Experiments}

\begin{table}[]
\centering
\normalsize
\caption{Analysis of Different Numbers of Local Epochs in Segmentation Task. (Communication Round/Local Epoch)}
\begin{tabular}{ccccc}
\hline
    & 1/300  & 2/150  & 4/75   & 5/60   \\ \hline
Dice & 88.02  & 88.21  & 87.52  & 88.00  \\
HD95 & 2.4560 & 2.3756 & 2.5918 & 2.5810 \\ \hline
\end{tabular}
\vspace{0.2cm}
\label{tab:com_seg}
\end{table}

In this subsection, We evaluate the impact of the hyperparameter $\mu$ on the performance, which is used to control the optimization step of calculating the correction model. The results are shown in Fig. \ref{fig_Ablation}, where the left and right vertical axes denote the PSNR and DSC values for imaging and segmentation tasks, respectively. We conduct experiments with $\mu$ from $1\times10^{-6}$ to 100. It can be observed that our method is not very sensitive to $\mu$, and our performance is better than other DCML methods even under the worst case. As suggested in Fig. \ref{fig_Ablation}, $1\times10^{-4}$ is chosen as the default selection in this paper.

The domain shift problem between clients may cause the global model to deviate from the global optimal solution after aggregation. We conduct experiments to sense the impact of the number of local training epochs and communication rounds on the performance. For all the cases, we set the numbers of training iterations equal, and the results of segmentation and restoration tasks are shown in Tabs. \ref{tab:com_seg} and \ref{tab:com_restor}, respectively. We notice that increasing the number of local training epochs would decrease the imaging performance, but the segmentation performance is not significantly affected. The domain gap in imaging task is greater than that in segmentation task since the scanner and scanning parameters may be different, which lead to a more serious model drift problem in imaging. The model drift is more serious as the number of local training epochs increases, which will lead the global model to deviate from the global optimal solution. As a result, we empirically decrease the number of local training epochs to avoid the above issues.

\section{Conclusions}
\label{sec:conclu}
Current U-shaped medical image networks have achieved impressive success without considering privacy issues. To simultaneously protect the data privacy of input, model parameters, output and label, we propose dynamic weight correction split federated learning (RoS-FL) for U-shaped medical image networks. Except for privacy protection, RoS-FL also has other important merits, such as low training time overhead and local computational resource. Meanwhile, we focus on the model drift problem in distributed learning, and propose dynamic weight correction strategy (DWCS) to correct the optimization solution and stabilize the training. Extensive experiments on different tasks demonstrate the effectiveness of our method. On the other side, although RoS-FL can achieve satisfactory performance on different domains, it ignores the test data belonging to an unseen domain. As a result, how to improve the performance for an unseen domain seems an interesting research field in our future work.

Another issue we must mention is that although the proposed RoS-FL achieves promising performance, one concern for our method is the risk of reconstructing raw images from shared feature maps. Related risks can be prevented by integrating privacy protecting techniques, such as differential privacy \cite{zhou2022differentially} and secure multi-party computation \cite{knott2021crypten, park2022evaluating}. Extending the proposed RoS-FL against inversion attacks by combining these technologies will be another possible research direction in the future.

\section*{Compliance with Ethical Standards}

This research study was conducted retrospectively using real clinical exams acquired at the University Hospital of Dijon and Mayo Clinic. Ethical approval was not required as confirmed by the license attached with the open access data.


%





\ifCLASSOPTIONcaptionsoff
  \newpage
\fi



%
\bibliographystyle{ieeetr}
\bibliography{ref}

%








\end{document}